# Utilizing Acceleration Measurements to Improve TDOA based Localization

Authors version

Marcin Kolakowski

Institute of Radioelectronics and Multimedia Technology, Warsaw University of Technology, Warsaw, Poland, contact: marcin.kolakowski@pw.edu.pl

**Additional information:**

The system for which the algorithm was originally intended is described in detail in:

J. Kolakowski, V. Djaja-Josko, M. Kolakowski; UWB Monitoring System for AAL Applications. Sensors 2017, 17, 2092. https://doi.org/10.3390/s17092092.

# Utilizing Acceleration Measurements to Improve TDOA based Localization

Marcin Kolakowski
Institute of Radioelectronics and Multimedia Technology
Warsaw University of Technology
Warsaw, Poland
m.kolakowski@ire.pw.edu.pl

*Abstract*—**In this paper localization using UWB positioning system and an inertial unit containing a single accelerometer is considered. The main part of the paper describes a novel algorithm for person localization. The algorithm is based on modified Extended Kalman Filter and utilizes TDOA (Time Difference of Arrival) results obtained from UWB system and results of acceleration measurement performed by the localized tag device. The proposed algorithm has been experimentally investigated through simulation and experiments. The results are included in the paper.**

## I. Introduction

Ultra-wideband positioning systems allow to localize objects in indoor environments with decimeter accuracy. However, such accuracy is hard to obtain, because it requires covering the whole analyzed space with system infrastructure, which may be expensive and difficult. Therefore other solutions like hybrid UWB-inertial localization systems are developed.

In these systems, in addition to conventional UWB radio transceivers, inertial measurement units (IMU) are used. Typical IMU contains a gyroscope tracking body's angular movement and an accelerometer. Systems based on UWB/IMU combination have been developed and allowed to achieve better positioning accuracy than in case of using inertial [1] or UWB system [2],[3] only.

One of the problems of the inertial measurement units is their level of power consumption, which is crucial in many applications. For example an IMU developed by Bosch Sensortec (BMI 160) has a typical current consumption about 0.85 mA [4]. In case of a single accelerometer (BMA 280) this consumption is considerably lower and is around 0.13 mA [5]. Therefore using both accelerometer and gyroscope is energy inefficient.

The possibility of gyroscope-free IMU construction has been investigated in [6]. Unfortunately recreating gyroscope functionality requires at least six accelerometers, which is equally demanding in case of power consumption.

In the paper an algorithm for person localization has been presented. The algorithm is intended for use in systems, where localized tag devices are equipped with an UWB transceiver and a single MEMS accelerometer. Such approach allows to combine some benefits of hybrid UWB-IMU systems with relatively low power consumption. The proposed algorithm combines TDOA (Time Difference of Arrival) results provided by system infrastructure with acceleration measurements conducted by the tag. In section II of the paper an architecture of the localization system is presented. Section III describes the localization algorithm. Simulation and experimental results are presented in section IV and V. Section VI concludes the paper.

## II. System Architecture

Architecture of the ultra-wideband system used in the paper is presented in Fig.1. The system consists of a localized tag device, system infrastructure comprising immobile synchronized anchors, Wi-Fi router and a system controller being a PC based application. The tag used in the system is a low energy device containing ultra-wideband transmitter and one MEMS accelerometer sensor.

In the presented system the tag is primarily localized based on Time Difference of Arrivals (TDOA) of transmitted signals. In order to obtain TDOA values each of the system anchors measures tag's packets times of arrivals (TOA). Besides being used for TOA measurements packets sent by the tags also act as data carriers for acceleration measurement results.

Measured TOA values and received accelerations are sent over the Wi-Fi link to the system controller, where TDOAs for selected anchor pairs are calculated. TDOA results are fused with measured acceleration using a novel algorithm and tag location is derived.

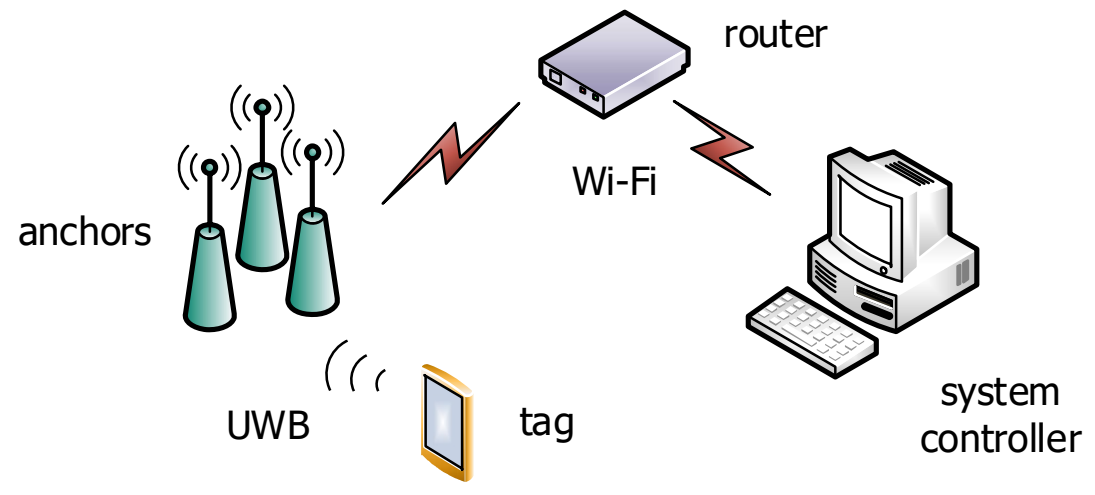

Fig. 1. Architecture of ultrawide-band localization system

## III. LOCALIZATION ALGORITHM

The proposed localization algorithm utilizes two types of measurement data: time difference of arrival measured by the system infrastructure and acceleration measured by the tag.

The accelerometer mounted on the tag conducts acceleration measurements in three perpendicular axes. Given that the tag contains no additional sensors such as a magnetometer or a gyroscope it is impossible to translate measured values to the specific directions in system's coordinates. Therefore only the value of resultant acceleration vector is used.

Persons localization is calculated using EKF (Extended Kalman Filter) based algorithm. In this algorithm, localized tag is modeled as a dynamic system, which state is described by its coordinates and velocity. The single iteration of basic EKF consists of two phases. In the first phase called "Time Update" current value of state vector is predicted based on state value obtained in the last EKF iteration and equations of motion. The predicted value is corrected based on measurement results during the second phase called "Measurement Update".

The algorithm proposed in the paper is a modification of Extended Kalman Filter [7]. Algorithm workflow scheme is presented in Fig.2.

The main difference in respect to the traditional EKF is the expansion of time update phase by adding additional prediction block. In the modified algorithm besides typical equations of motion based prediction another parallel prediction using step analysis is performed. Step analysis is conducted using acceleration measurements and previously determined persons locations. Tag measured accelerations are used to detect steps, determine step frequency $f_s$ and step length $L_s$. Recorded persons positions are used to determine heading $\theta$. State vector values obtained in both prediction blocks are fused and then updated with TDOA measurement results.

### A. Heading estimation

Heading of a moving person is derived from previously calculated tag localizations. An exemplary set of points and estimated movement directions are presented in Fig.3.

Movement direction is estimated using linear regression performed for a set of a few previously calculated positions. The proposed method assumes that the localized person was moving along a straight line during the time those samples were collected. Such assumption is justified for use in UWB localization systems, which are known for their relatively high position refresh rate. For example in the system which allows to calculate user's position ten times per second and direction is determined based on five previous locations, heading would be calculated for a short period of time equal to 0.5s. Given that person walking at normal pace usually does not rapidly change movement direction such approximation seems acceptable.

Linear regression allows to determine parameters of a line best fitting trajectory alongside which the localized person was moving during few previous moments. In the system, heading is described by θ, which is an angle between the derived line and Y axis.

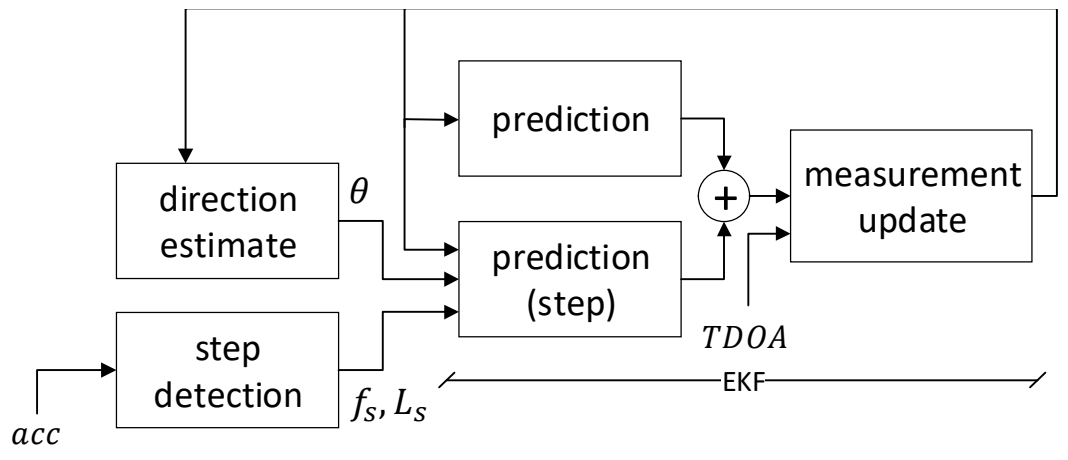

Fig. 2. Algorithm worklflow diagram

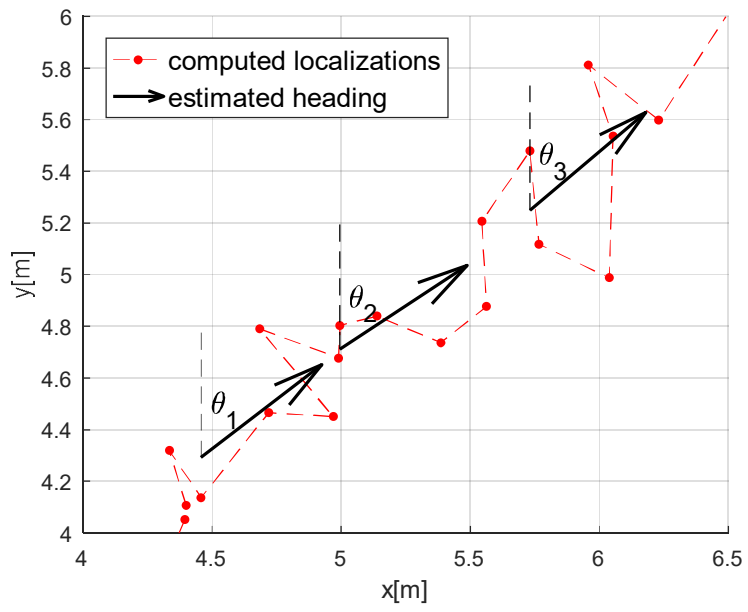

Fig. 3. Localization results and estimated movement directions. The direction is described by angle $\theta$.

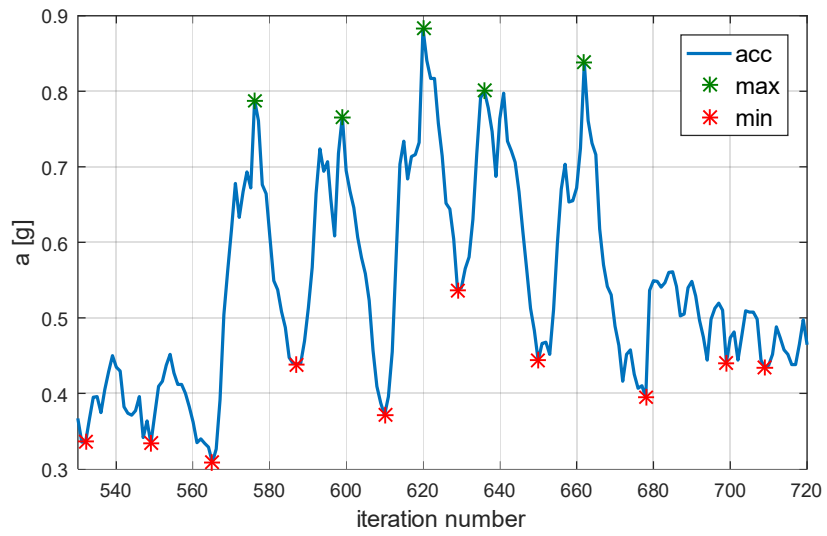

Fig. 4. Smoothed acceleration values measured during a five-step sequence, green and red dots indicate detected peaks and valleys respectively

### B. Step detection

The accelerometer sensor mounted on the tag device measures acceleration with frequency which is five times higher than localization system refresh rate. Therefore each of the UWB packets sent by the tag carries five sets of tri-axial acceleration samples. As mentioned earlier, the algorithm uses value of the resultant acceleration which is additionally smoothed using moving average. Acceleration measurement results are used for localized persons step and stop detection. An exemplary measured acceleration graph is shown in Fig.4.

Localization algorithm works with a delay of a few samples, so that a more advanced than simple threshold step detection can be performed. Step detection procedure begins with locating peaks and valleys in the signal. In order to prevent false step detections, only peaks above some fixed value are considered. If a peak is located, the algorithm analyzes the preceding samples. The sample for which the detection threshold was exceeded is treated as the moment of step's start. Persons step end is determined by the position of the first valley following the peak. The times corresponding the selected samples are used to derive step frequency and with addition of localization results length of persons step.

## C. *Position calculation*

Tag localization is calculated using a modified Extended Kalman Filter algorithm. In this algorithm a tag is treated as a dynamic system, which state at moment $x_k$ is described by a vector containing its x, y coordinates and speed vector components (1).

$$x_k = [x\ y\ v_x v_y]^T \tag{1}$$

In the proposed algorithm, during the time update phase, two separate predictions are performed. The first prediction is state prediction based on a previous state values and equations of motion:

$$\hat{x}^{(1)}_{k(-)} = F\hat{x}_{k-1(+)} \tag{2}$$

$$P^{(1)}_{k(-)} = FP_{k-1(+)}F^T+Q^{(1)}_k \tag{3}$$

where $\hat{x}^{(1)}_{k(-)}$ and $\hat{x}_{k-1(+)}$ are predicted state vector value and state vector obtained in the previous filter iteration, $P_{k(1-)}$ and $P_{k-1(+)}$ are their covariance matrices and $F$ is state transition matrix. $Q^{(1)}_k$ is covariance matrix of process noise and was chosen according to Discrete White Noise Acceleration Model (DWNA) [8].

The second conducted prediction is based on localized persons step analysis. When the step was detected and the person is moving the following equations are used:

$$\hat{x}^{(2)}_{k(-)} = \hat{x}_{k-1(+)}+[\sin\theta \quad \cos\theta \quad 0 \quad 0]^T f_s L_s \Delta T \tag{4}$$

$$P^{(2)}_{k(-)} = P_{k-1(+)}+Q^{(2)}_k \tag{5}$$

where $\theta$ is movement direction, $f_s$ and $L_s$ are step frequency and length and $\Delta T$ is system refresh period. When the person is not moving prediction equations are as follows:

$$\hat{x}^{(2)}_{k(-)} = F_2\hat{x}_{k-1(+)} \tag{6}$$

$$P^{(2)}_{k(-)} = F_2P_{k-1(+)}F_2^T+Q^{(2)}_k \tag{7}$$

where $F_2$ is a modified dynamics matrix, which takes into account the knowledge that the localized person stopped moving. Both predictions are combined in a way known from traditional Kalman Filter.

$$K_k^{(step)} = P^{(1)}_{k(-)}\left(P^{(1)}_{k(-)}+P^{(2)}_{k(-)}\right)^{-1} \tag{8}$$

$$\hat{x}_{k(-)} = \hat{x}^{(1)}_{k(-)}+K_k^{(step)}\left(\hat{x}^{(2)}_k-\hat{x}^{(1)}_{k(-)}\right) \tag{9}$$

$$P_{k(-)} = \left(I-K_k^{(step)}\right)P^{(1)}_{k(-)} \tag{10}$$

The resulting value of state vector is then updated with TDOA measurements performed by the system infrastructure (11-14).

$$h_k(x_k)=[TDOA_1(x_k) \quad \cdots \quad TDOA_n(x_k)] \tag{11}$$

$$K_k=P_{k(-)}H_k^T\left(H_kP_{k(-)}H_k^T+R_k\right)^{-1} \tag{12}$$

$$\hat{x}_{k(+)}=\hat{x}_{k(-)}+K_k\left(z_k-h_k\left(\hat{x}_{k(-)}\right)\right) \tag{13}$$

$$P_{k(+)}=\left(I-K_kH_k^T\right)P_{k(-)} \tag{14}$$

where $z_k$ is a measurement vector containing TDOA measurement results, $h_k(x_k)$ is the measurement function, which is used to calculate TDOA corresponding to those in $z_k$ based on predicted tag position, $H_k$ matrix is that function's linearization. The size of measurement vector is not constant and depends on the number of available TDOA results. The presented algorithm allows to locate the person in two dimensions but it can be easily extended and adapted for three-dimensional localization.

# IV. Simulation Results

The presented algorithm was tested during simulation step. The simulations were performed in Matlab environment. The goal of the simulation was to verify the algorithm and compare it with algorithms utilizing solely TDOA measurement results.

Simulations were conducted for a system consisting of four anchor nodes placed on the walls of a large rectangular room. In the simulation scenario a localized person was moving alongside a pentagon shaped trajectory. It was assumed that the person was wearing the tag attached to his belt, which led to blocking direct propagation path between the tag and some parts of system infrastructure. This effect was taken into account by adding a few ns delay to TOAs measured by obstructed anchors. It was assumed that the anchors measured TOA with standard deviation equal to 0.3 ns.

Tag localization was calculated using three different algorithms: Least-Squares estimator, EKF using TDOA and the algorithm presented in the paper. Exemplary simulation results are presented in Fig.5. The accuracy of the algorithms can be compared based on empirical CDF curves plotted for localization error (Fig.6).

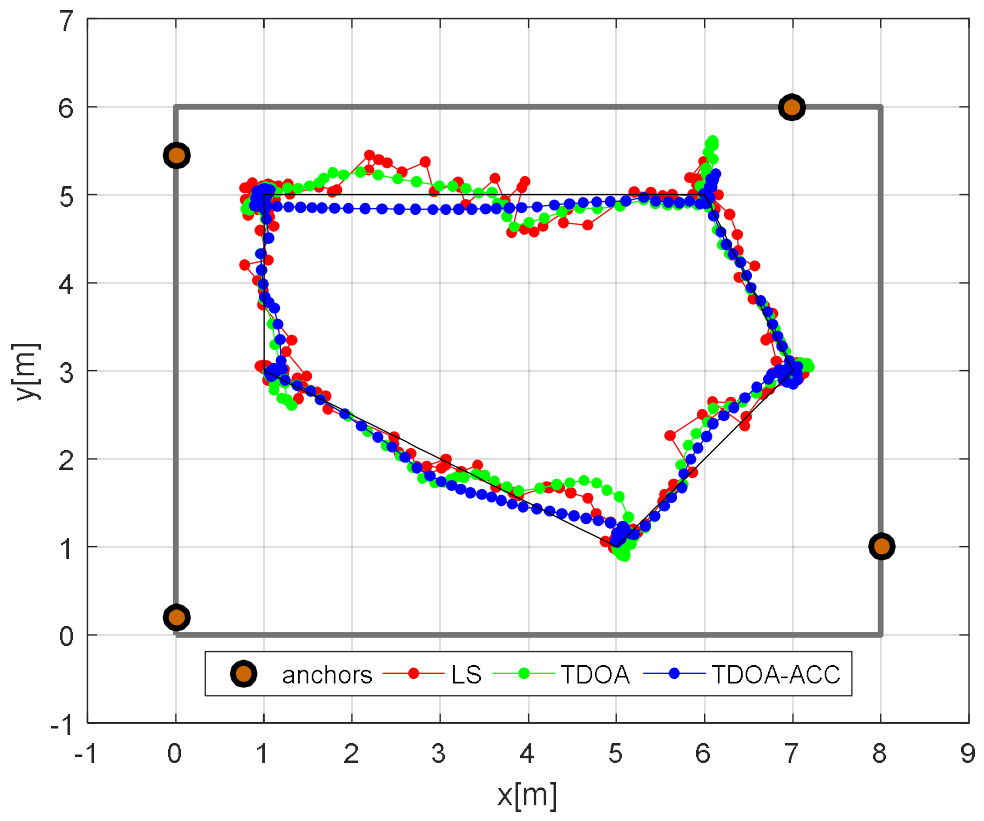

Fig. 5. Simulation results

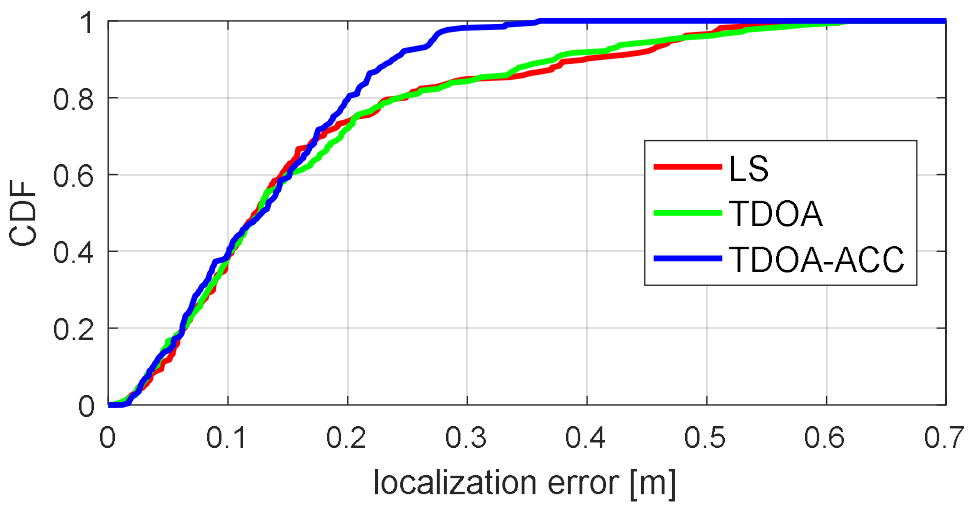

Fig. 6. CDF for trajectory error

In case of LS and EKF using TDOA algorithms blocking the direct path between the tag and system infrastructure results in deviations from the reference trajectory. The use of novel algorithm allows to suppress these effects. The maximum error obtained with the proposed algorithm is about 35 cm, whereas for LS and EKF it is higher and is about 60 cm.

## V. Experimental Results

The presented algorithm was experimentally tested. The tests were performed in in the laboratory room of size approx. 6 by 6 meters. The test setup included UWB positioning system developed within NITICS project [9]. The infrastructure of the system consisted of four anchors placed on rooms walls. In this room, such number of anchors and their placement is sufficient to provide the user with reliable and accurate localization. In case of larger or more complicated areas e.g. fully furnished apartments, the number of anchors should be higher.

The tag device was attached to localized persons belt. During the experiments the person was walking along the trajectory consisting of three straight lines. Localization was performed using three algorithms: Least-Squares estimator, EKF using TDOA and the algorithm combining TDOA with acceleration results. Localization results are presented in Fig. 7.

The results obtained with the novel algorithm are a bit more accurate than those for LS and TDOA. The advantages of algorithm combining TDOA and acceleration are visible in two situations.

Firstly additional prediction based on movement direction and step analysis helps to reduce negative effects of NLOS (Non Line of Sight) conditions caused by persons body blocking the direct visibility between the tag and some parts of the infrastructure. These negative effects can be observed in case of LS and TDOA algorithms, where deviations from straight line trajectory are present. In case of the novel algorithm the computed trajectory is closer to the original straight line.

Secondly, algorithms awareness of whether the localized person is walking or not is useful when the person is turning around at the end of the path. In this situation work conditions change rapidly from LOS to NLOS for all anchors. For LS and TDOA it results in deriving positions are forming a loop which is too large for a normal turnaround. The novel algorithm helps to suppress this effect.

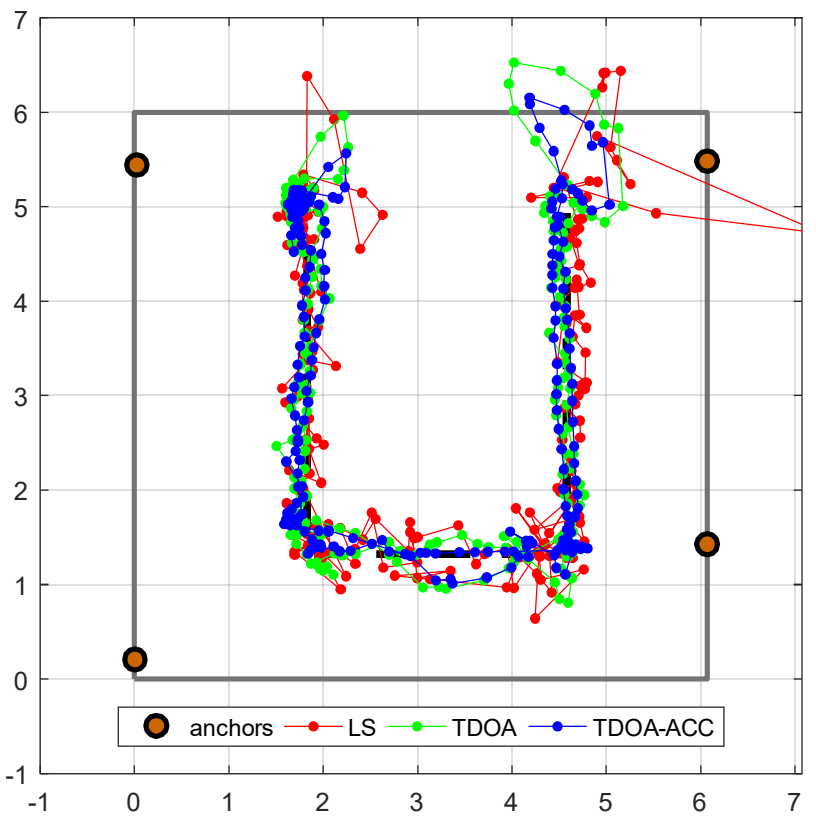

Fig. 7. Visualization of located points

## VI. Conclusions

In the paper a novel algorithm for use in ultra-wideband positioning systems is presented. The algorithm allows to combine TDOA results and acceleration measured by the tag. The use of only an UWB radio module and a single accelerometer allows to keep tag's power usage on low levels, which prolongs battery longevity.

Experimental results have shown that the novel algorithm allows to localize walking persons with higher accuracy than widely used Least-Squares estimator and simple EKF using only TDOA results.

The presented algorithm can be modified and upgraded in terms of step detection and heading estimation techniques.